\documentclass{aa}
\usepackage{natbib,psfig,graphicx,ulem,epsfig,psfig}
\usepackage{txfonts}
\usepackage{soul,color}
\usepackage{longtable}
\sethlcolor{white}

\def\mathnew{\mathsurround=0pt}
\def\simov#1#2{\lower .5pt\vbox{\baselineskip0pt \lineskip-.5pt
\ialign{$\mathnew#1\hfil##\hfil$\crcr#2\crcr\sim\crcr}}}

\begin{document}

\title{A search for faint low surface brightness galaxies in the relaxed
  cluster Abell~496 \thanks{Based on observations obtained with
    MegaPrime/MegaCam, a joint project of CFHT and CEA/DAPNIA, at the
    Canada-France-Hawaii Telescope (CFHT) which is operated by the
    National Research Council (NRC) of Canada, the Institut National
    des Sciences de l'Univers of the Centre National de la Recherche
    Scientifique (CNRS) of France, and the University of Hawaii. The
    data processing were performed by the TERAPIX Data Centre.}}

\author{
M.~P.~Ulmer \inst{1,2} \and
C.~Adami \inst{1} \and
F.~Durret \inst{3,4} \and
O.~Ilbert \inst{1} \and
L.~Guennou \inst{1} 
}

\institute{
LAM, P\^ole de l'Etoile Site de Ch\^ateau-Gombert,
38 rue Fr\'ed\'eric Joliot-Curie,
13388 Marseille Cedex 13, France
\and
Department of Physics and Astronomy, Northwestern University, 2131 
Sheridan Road, Evanston IL 60208-2900, USA
\and
UPMC Universit\'e Paris 06, UMR~7095, Institut d'Astrophysique de Paris, 
F-75014, Paris, France
\and
CNRS, UMR~7095, Institut d'Astrophysique de Paris, F-75014, Paris, France
}

\date{Accepted . Received ; Draft printed: \today}

\titlerunning{Faint low surface brightness galaxies in Abell 496}

\abstract
{Cluster faint low surface brightness galaxies (fLSBs) are difficult to
  observe.  Consequently, their origin, physical properties and number
  density are not well known. After a first search for fLSBs in the
  highly substructured Coma cluster, we present 
  here a search for fLSBs in the nearly relaxed Abell~496 cluster.}
{Abell~496 appears to be a much more relaxed cluster than Coma, but still 
  embedded
  in a large scale filament of galaxies. Our aim is to compare the
  properties of fLSBs in these two very different clusters, to search
  for environmental effects.}
{Based on deep CFHT/Megacam images in the $u^*$, $g'$, $r'$ and $i'$
  bands, we selected galaxies with $r'>21$ and $\mu_{\rm{r'}}> 24$ mag
  arcsec$^{-2}$. We estimated photometric redshifts for all these
  galaxies and kept the 142 fLSBs with photo$-z<0.2$.}
{In a $g'-i'$ versus $i'$ color-magnitude diagram, we find that a
  large part of these fLSBs follow the red sequence (RS) of brighter 
  galaxies. The fLSBs within $\pm 1\sigma$ of the RS show a homogeneous
  spatial distribution, while those above the RS appear to be
  concentrated along the large scale filament of galaxies.}
{These properties are interpreted as agreeing with the idea that RS fLSBs
  are formed in groups prior to cluster assembly. The  formation of red fLSBs
  could be related to infalling galaxies.}


\maketitle

\section{Introduction}

Faint low surface brightness galaxies (fLSBs hereafter) remain a
poorly known class of galaxies, though they are interesting objects
for several reasons, as already discussed in detail by Adami et
al. (2009a).  We define fLSBs as galaxies with a central surface
  brightness fainter than $\mu_{\rm{r'}}$ = 24 mag arcsec$^{-2}$ and a
  total magnitude $r'>21$, to be consistent with Adami et al. 2006,
  hereafter ASU06.  Briefly: fLSBs could account for part of the
missing low luminosity structures predicted by CDM models of
hierarchical structure formation (White \& Rees 1978), in particular
since they appear dominated by dark matter (e.g. McGaugh et al. 2001,
de Blok et al. 2001). CDM models predict the existence of low
luminosity galaxies in all environments, but fLSBs seem to be present
in higher numbers in clusters than in the field (see e.g. Sabatini et
al. 2005, ASU06, and references therein).

Many fLSBs are fainter than the night sky and clearly extend toward
fainter brightnesses than predicted by the Freeman law (1970), as
shown for example by Bothun et al. (1997).  Due to their extreme
faintness both in terms of surface brightness and of total magnitude,
fLSBs are therefore very difficult to detect, hence their origin,
physical properties and number density are not well known in a
statistical way over a large number of clusters, despite numerous
studies (e.g. Binggeli et al. 1985; Schombert et al. 1992; Bothun et
al. 1993; Bernstein et al. 1995; Impey et al. 1996; Sprayberry et
al. 1996; Ulmer et al. 1996; Impey \& Bothun 1997; O'Neil et al. 1997;
Kuzio de Naray et al. 2004).

In order to increase the number of fLSBs detected in clusters, our
team has searched for Coma cluster fLSBs in the total magnitude versus
central surface brightness space (ASU06, Adami et al. 2009a) and found
for example that these objects tended to be more concentrated in
several areas (not always central).  Furthermore, based on their
position in the (B$-$R) versus R plane, we found that we could
identify three distinct types of fLSBs. Those that fall on the color
magnitude relation extrapolated from the bright normal galaxy
population we called $sequence$ fLSBs. We interpreted $sequence$ fLSBs
as galaxies that formed in small groups prior to the cluster
assembly. Then we interpreted the reddest fLSBs as faint stripped
ellipticals and the blue fLSBs as galaxies made of material stripped
from spiral infalling galaxies.  However, the Coma cluster is highly
substructured (e.g. Adami et al. 2005) and we do not know how
substructure could affect the spatial distribution of the fLSB
population.  We therefore decided to analyze in the same way the
distribution and properties of fLSBs in a more relaxed cluster where
substructures will not complicate the picture.

Abell~496 is one of the rare nearby nearly relaxed clusters (see
e.g. Durret et al. 2000). Bou\'e et al., (2008) reported the
detailed analysis of the galaxy luminosity functions of Abell~496,
based on deep CFHT Megacam images in four bands which are ideal to
search for fLSBs. They confirmed that this cluster appears very relaxed,
with no particular structure at the cluster scale, though at
larger scale an extended filament of galaxies with redshifts close to
that of Abell~496 was found to spread from the north-west to the
south-east of the cluster (see Fig.~10 in Bou\'e et al. 2008).

The mean heliocentric velocity of Abell~496 is cz=$9885 \, \rm km \,
s^{-1}$, corresponding to a redshift $z = 0.0329$, its distance
modulus is 35.69, and the scale is 0.666 kpc arcsec$^{-1}$, assuming
H$_0=72$~km~s$^{-1}$~Mpc$^{-1}$, $\Omega_{\rm M}=0.3$ and
$\Omega_{\Lambda}=0.7$.  It has an angular virial radius of
$0.77^\circ$ (1.85~Mpc), obtained by extrapolating the radius of
overdensity 500 (Markevitch et al.  1999), measured relative to the
critical density of the Universe to the radius of overdensity 100.  We
will give magnitudes in the AB system.

The paper is organized as follows.  The data and method to search for
fLSBs are described in Section~2. Results concerning the
color-magnitude relation, spatial distribution and luminosity function
of fLSBs are presented in Section~3 and discussed in Section~4. We give
in the Appendix the list of the 142 fLSBs with photo$-z<0.2$ as well as
the images in the four bands and the surface brightness profile for
one of them.

\section{The data and method}

\subsection{The optical data}

This work is based on deep images obtained at the CFHT with the
Megaprime/Megacam camera (program 03BF12, P.I. V. Cayatte) in the four
bands $u^*,\ g',\ r',\ i'$ already described in detail by Bou\'e et
al. (2008).  The images are centered on the cluster centre as
  defined by NED: J2000.0 equatorial coordinates
  04$^h$33$^{mn}$37.1$^s$,$-$13$^\circ$14$'$46$''$. They were reduced by
the TERAPIX pipeline. Since simple detection with SExtractor (Bertin
$\&$ Arnouts 1996) is not always sufficient to measure fLSB magnitudes
unambiguously, we applied the same elaborate technique as in ASU06,
which is briefly described below.

\subsection{The method to search for fLSBs}

In order to make the comparison with the fLSBs in Coma
straightforward, we detected fLSBs with the same method as described
in ASU06.  In brief, we started with a catalog produced by SExtractor
from the Abell~496 CFHT Megacam images.  Then, since our fLSB
dedicated software could not be applied to such large images, we
divided each image (and the corresponding catalogue) in 25 subimages,
each $0.2\times0.2$~deg$^2$.

The first cut was to eliminate bright objects (total magnitudes $r' <
21$) from our analysis in order to be consistent with the ASU06
selection process.  This selection criterion is based on the fact that
part of the cluster fLSBs could be tidal dwarf galaxies (see
ASU06). Tidal dwarf galaxies have masses as low as 10$^7$ or
10$^8~M_\odot$ (Bournaud et al. 2003), and as shown in ASU06 this
translates to magnitudes fainter than $r'\sim$21.  Each of the 25
subimages was then examined visually, in order to note areas around
diffraction spikes and between CCDs, and all SExtractor objects in
these areas were removed from further analysis.

As we had images in four bands and our software was designed to
process only two bands at once, we first considered the $r'$ and $u^*$
bands, in order to encompass the 4000~\AA\ break at the redshift of
Abell~496 ($z = 0.0329$). We then ran a three--step iterative
selection process on each of the 25 subimages and for the $r'$ and
$u^*$ bands to generate a primary data set.

First, we fit a Gaussian form plus a constant background to the
linear-scale surface brightness profiles on the images, as in ASU06.
Although fLSBs have exponential surface brightness profiles, Ulmer et
al. (1996) found that fLSB selection based on exponential profiles
generates a large number of false candidates in rich environments, due
to the proximity of neighboring objects. Instead of using exponential
profiles, ASU06 therefore selected fLSBs by $\chi ^2$-fitting of
Gaussian curves to the radial surface brightness profiles of
fLSBs. This does not mean that an exponential is not the proper form
of fLSB profile. Rather, the Gaussian profile is the result of the
intrinsic (exponential) shape convolved with instrumental effects (the
PSF, due to seeing, had typical values between 0.4 and
0.6~arcsec). Initially, we let the radial profiles extend to a maximum
radius $\theta_{\rm{max}}= 2.5$~arcsec from the center of each object,
which, as determined by visual inspection, encompasses the entire
range of fLSB sizes.

Second, we selected initial fLSB candidates with radius greater than
0.6~arcsec.  The radius is defined here as the $\sigma$ and not as the
FWHM of the profile (FWHM = 2.35 $\sigma$). The size threshold was
chosen above the seeing radius in order to limit contamination by
globular clusters which at the distance of Abell~496, appear as point
sources. The $r'$ central surface brightness was chosen fainter than
$\mu_{\rm{r'}}$ = 24 mag arcsec$^{-2}$ to be consistent with ASU06.

Third, we optimized the final value of $\theta_{\rm{max}}$ for all the
selected candidates to ensure that none of their surface brightness
profiles were contaminated by surrounding objects. This process is
explained in more detail in ASU06.  The optimized $\theta_{\rm{max}}$
for each candidate was determined by visual inspection. We then
repeated the two previous steps. After inspecting all candidates
visually we selected as final fLSBs the candidates that yielded an
acceptable Gaussian fit to a distance of $\theta_{\rm{max}}$ (see also
ASU06 for more details). By ``acceptable'' we mean that the
probability of finding a better fit (by changing the parameters) is
smaller than 10\%.  An example is shown in Fig.~\ref{fig:prof}.

The convergence/non-convergence was done as follows: if the
chi-squared changed by less than 0.1\% within 20 iterations, this was
called convergence. If the chi-squared failed to decrease by less than
0.1\% in 20 iterations or if the chi-squared actually grew without
bound, then this was called non-convergence.

The final data set was defined requiring a good (i.e. converging) fit
for both the $r'$ and $u^*$ bands simultaneously. We then computed
magnitudes for this sample in the $g'$ and $i'$ images.  The automated
analysis produced valid Gaussian fits most of the time. For objects
with a non converging process (for example only 3 cases for the $g'$
band), we calculated the missing $i'$ and $g'$ band magnitudes by
comparing the SExtractor magnitudes with the results from our
dedicated code for the fLSBs with a converging process. Then we
applied the relation deduced in this way to the fLSBs with a non
converging process.

\subsection{Cluster membership of the fLSBs}

\begin{figure} 
\centering \mbox{\psfig{figure=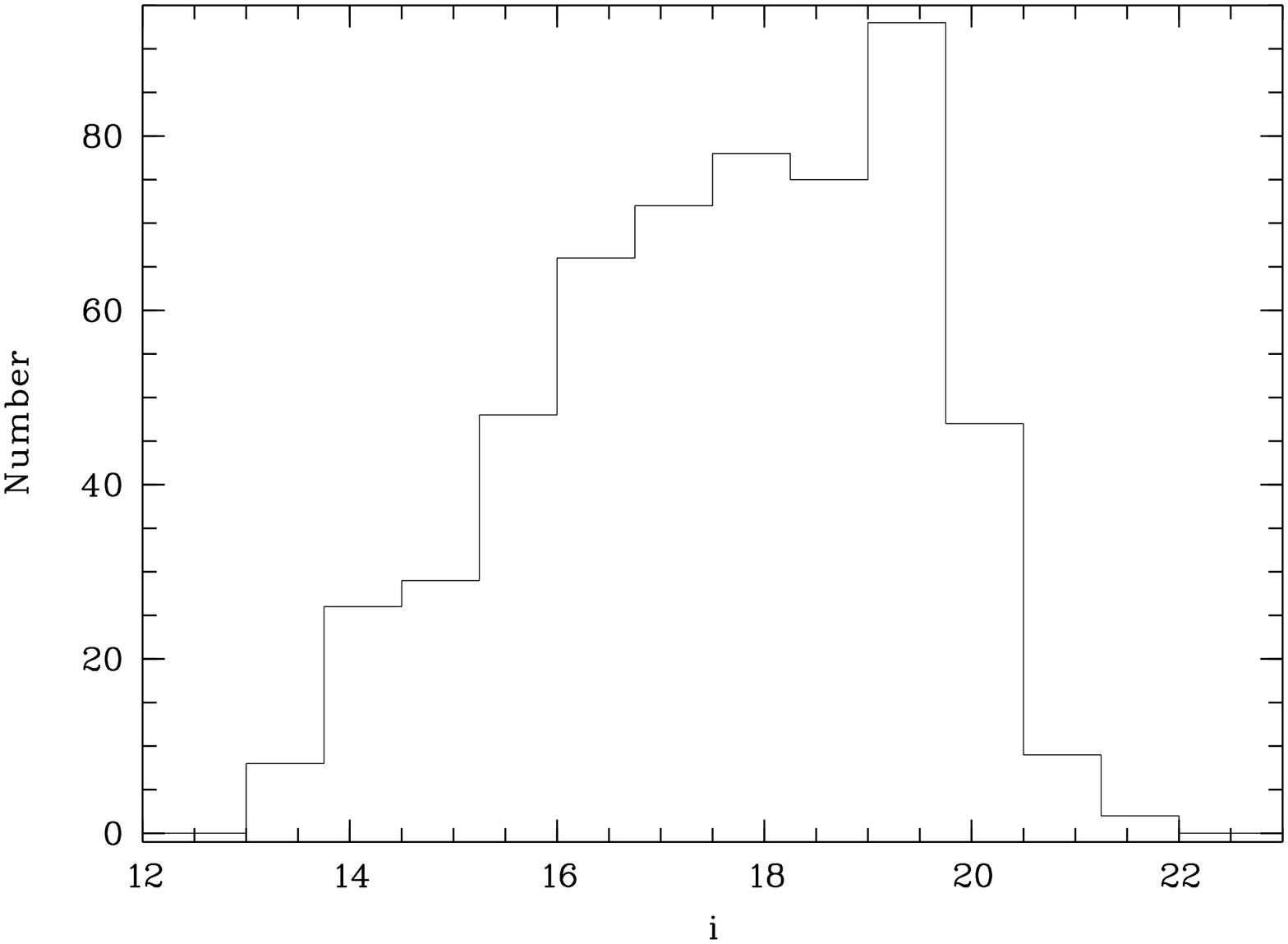,width=5.0truecm,angle=270}}
\centering \mbox{\psfig{figure=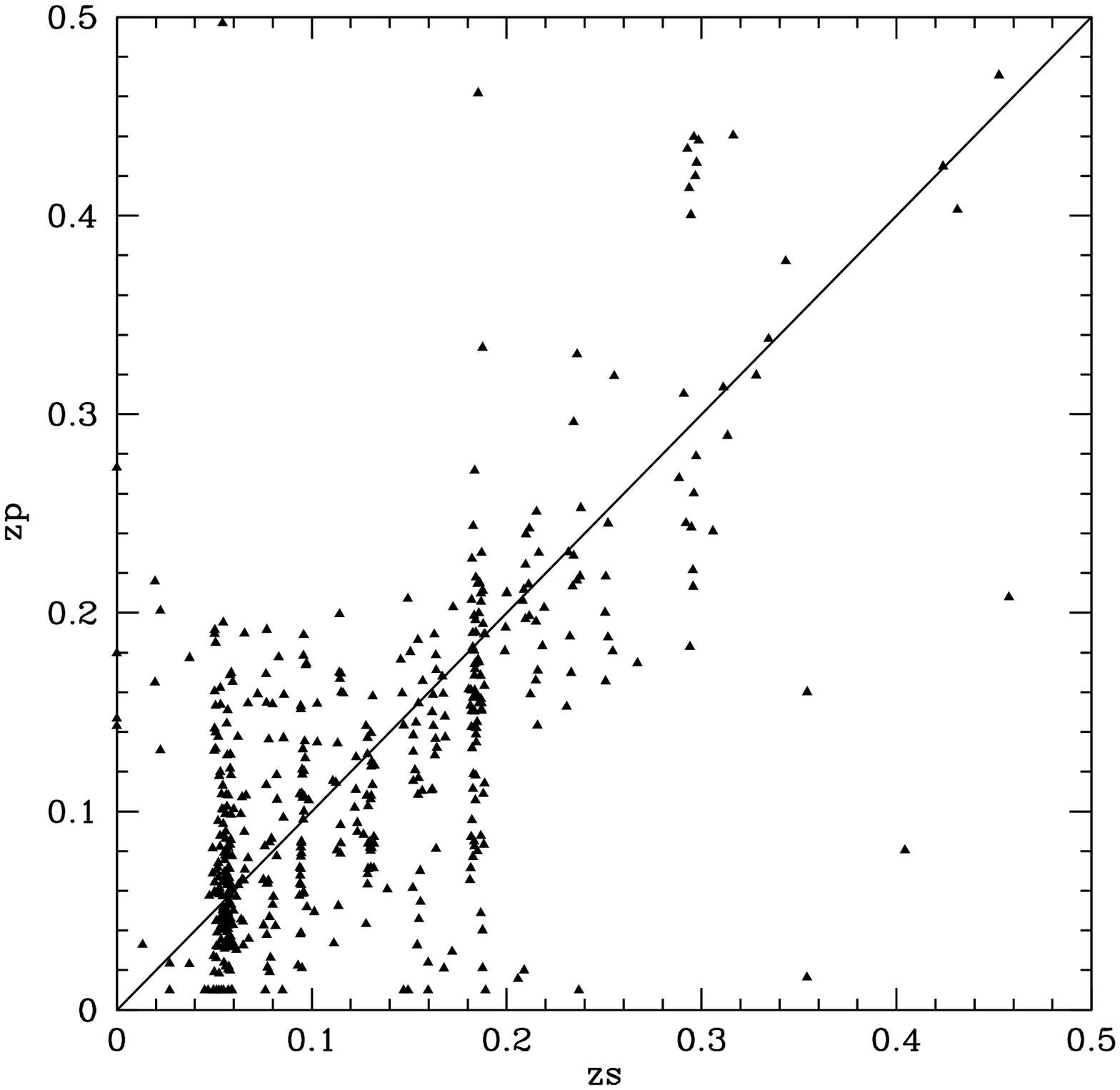,width=6.0truecm,angle=0}}
\caption[]{Upper figure: $i'$ band magnitude histogram of our
  spectroscopic sample. Lower figure: photometric versus spectroscopic
  redshifts. }
\label{fig:histoi}
\end{figure}

Independently, we calculated the photometric redshifts (hereafter
photo$-z$s) for all the galaxies detected in the images, based on the
SExtractor magnitudes in the four bands by applying the LePhare
software (Ilbert et al. 2006). The zero point of each band was
adjusted using a spectroscopic catalog of 596 galaxies brighter than
$i'\sim$19.5.  Fig.~\ref{fig:histoi} clearly shows that we can
efficiently discriminate between z$\ge$0.2 and z$<$0.2, as most
objects with photo$-z$s $<$0.2 also have spectroscopic redshifts
$<$0.2. This redshift value of 0.2 was also found to be optimal by
Adami et al. (2008) with similar data.

We thus produced the two following fLSB samples: (a) fLSBs
with a photo$-z$ below 0.2 (142 galaxies); (b) fLSBs with a
photo$-z$ above 0.2 (783 galaxies).

The goal being to study cluster galaxies, we must estimate the
contamination of the sample of 142 fLSBs by non cluster galaxies. From
Fig.~\ref{fig:histoi}, the expected contamination of the z$<$0.2
redshift interval by z$\ge$0.2 fLSBs is of the order of 5\% (7
galaxies among the 142).

We must also estimate how many fLSBs at z$<$0.2 are not part of
Abell~496. Since Abell~496 is at z$\sim$0.033, there is a
non-negligible cosmological volume behind the cluster, as the
photometric redshift technique is not accurate enough to distinguish
between a cluster member and a z$<$0.2 non-cluster galaxy.

One method to estimate the number of z$<$0.2 non cluster member fLSBs
is to estimate the volume density of field fLSBs. This is not a
trivial task as our selection function is quite specific and is not
reproduced by most literature studies. However we can take advantage
of the deep spectroscopic follow up of the Coma cluster of Adami et
al.  (2009b), where a spectroscopic redshift was successfully measured
for eleven fLSBs along the line of sight to Coma, selected exactly in
the same way as in the present paper. These fLSBs all had a
photometric redshift (computed in the same way as here) lower than 0.2
and were brighter than I=22.7. Four of these eleven galaxies proved
not to be part of the Coma cluster, though they were at redshifts
lower than 0.2. This gives 21 galaxies per deg$^2$ at photo$-z <$0.2
and I$\le$22.7 which are not cluster members. Along the line of sight
to Abell~496, we detected 122 fLSBs brighter than i$'=23.2$
(equivalent to I=22.7, see Fukugita et al. 1995) in a 1~deg$^2$ field,
so 21 of these should therefore not be part of the
cluster. Extrapolating this number to the complete magnitude range, 24
of the 142 detected fLSBs at photo$-z < $0.2 are expected not to be
members of Abell~496. 

Another method to estimate how many fLSBs at z$<$0.2 are not part of
Abell~496 is based on the assumption that cluster fLSBs follow a King
number density distribution. This method results in $\sim$24$\pm$22
fLSBs at z$\le$0.2 being non cluster members (also see section 3.2),
in agreement with the previous estimate, though with a large error.

We therefore conclude that among our 142 photo$-z <$0.2 fLSBs, about
30 galaxies (24 effectively located at z$<0.2$ but not in the cluster,
plus 7 at z$>0.2$ but classified as being at z$<0.2$), or 21\%, may not
be part of Abell~496.

\section{Results}

The list of our 142 $z<$0.2 candidate fLSBs is given in
Tables~\ref{tab:liste}, ~\ref{tab:liste2}, and ~\ref{tab:liste3} with
their positions, four band magnitudes as measured by the present
process, and photometric redshifts.

\subsection{Color-magnitude relation}  
\label{sec:cmr}

\begin{figure} 
\centering \mbox{\psfig{figure=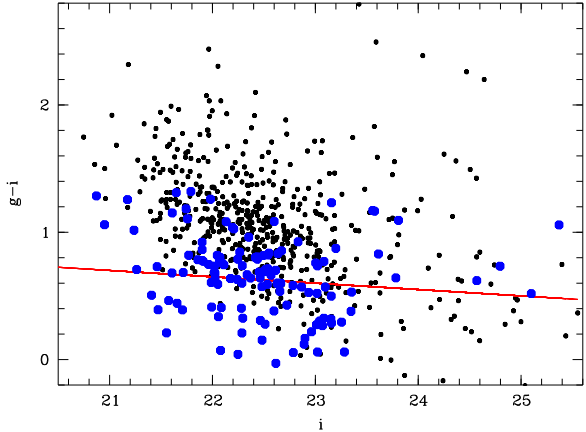,width=6.0truecm,angle=-90}}
\caption[]{Color--magnitude relation for all the fLSBs in the direction
  of Abell~496 (in black), and for the 142 galaxies for which the
  photo$-z$ is less than 0.2 (in blue). The red line corresponds to the
  red sequence for bright galaxies (see text). }
\label{fig:cmrbig}
\end{figure}

\begin{figure} 
\centering \mbox{\psfig{figure=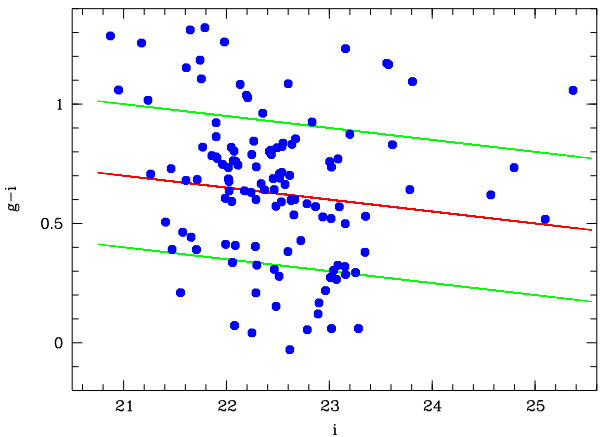,width=6.0truecm,angle=-90}}
\caption[]{Zoom on the color--magnitude relation for the fLSBs in the
  direction of Abell~496 with photo$-z<0.2$ (142 objects). The red
  line corresponds to the red sequence for bright galaxies (see text).
  The green lines delineate the region above and below the red
  sequence, that we will respectively call $red$ and $blue$ fLSBs.  The
  separation between these two lines corresponds to the approximate $1
  \sigma$ loci, $\pm 0.29$~mag on either side of the red sequence.  }
\label{fig:cmrsmall}
\end{figure}

The $g'-i'$ versus $i'$ color-magnitude relation obtained for our
fLSBs is plotted in Fig.~\ref{fig:cmrbig}, together with the
color-magnitude relation found by Bou\'e et al. (2008) for the
``normal'' galaxies of Abell~496.  The red sequence defined for the
galaxies belonging to Abell~496 was computed by Bou\'{e} et al. (2008)
for galaxies brighter than $i'\sim 21$ to be: $g'-i'=-0.05i'+1.75$.
We can see in Fig.~\ref{fig:cmrsmall} that most of the fLSBs with
  photo$-z<0.2$ fall close to the color-magnitude relation defined
    by brighter normal galaxies from Bou\'e et al. (2008), though
    there is a non-negligible scatter.

In contrast, the fLSBs with photo$-z$ $> 0.2$ are mostly
located above the color-magnitude relation, suggesting that they are
mostly redder and therefore background objects. This is not
surprising, as the photometric redshift selection is primarily
based on colors and therefore defines relatively blue colors at low
redshift and relatively red colors for higher redshift.

We show in Fig.~\ref{fig:cmrsmall} a zoom of the color-magnitude
relation for the 142 fLSBs with photo$-z <0.2$. We can define three
subsamples: the $sequence$ fLSBs (within $\pm 1\sigma$, or $\pm
0.29$~mag from the red sequence), the $blue$ fLSBs (more than
$1\sigma$ below the red sequence), and the $red$ fLSBs (more than
$1\sigma$ above the red sequence).  This classification is similar to
that already proposed for fLSBs in Coma (ASU06), suggesting that a
large fraction (here about 2/3) of fLSBs follows an evolutionary path
comparable to that of normal ellipticals in clusters. We will discuss
this result in more detail in Section~\ref{sec:discu}.

\subsection{Spatial distribution of the fLSBs and cluster substructure}

\begin{figure} 
\centering \mbox{\psfig{figure=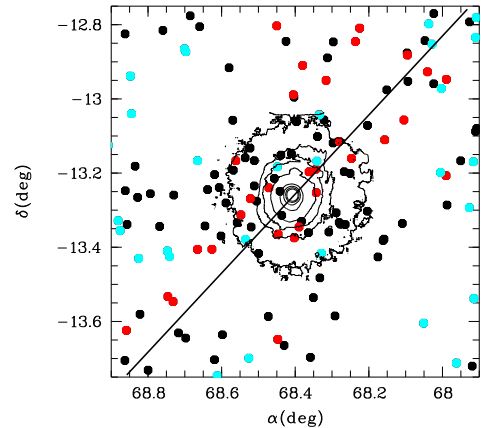,width=6.0truecm,angle=-90}}
\caption[]{Positions of the fLSBs with photo$-z<0.2$. Black, cyan and
  red points correspond to fLSBs within $\pm 1\sigma$ of the red
  sequence, below, and above this interval respectively. The black
  line indicates the direction of the very large scale filament of
  galaxies found by Bou\'e et al. (2008) -- see their Figure~10. 
  The contours correspond to the X-ray emission from the XMM-Newton
  EPIC MOS1 image (Lagan\'a et al. 2008). }
\label{fig:ad}
\end{figure}

A bi-dimensionnal Kolmogorov Smirnov (KS hereafter) test shows that
the $\alpha$,\ $\delta$ spatial distribution of fLSBs at z$<$0.2 is
different at the 92$\%$ level from a uniform distribution; the same KS
test shows that the spatial distributions of the z$<$0.2 and
z$\geq$0.2 fLSBs are different at the 99.9$\%$ level. The fLSBs with a
high probability of belonging to Abell~496 are therefore not as
uniformly distributed throughout the cluster as the galaxies likely to
be non-cluster members.

The $\alpha$,\ $\delta$ spatial distributions of the photo$-z<0.2$
$sequence$, $red$, and $blue$ fLSBs shown in Fig.~\ref{fig:ad} are
also different. The distribution of $blue$ fLSBs is different from a
uniform spatial distribution only with a probability of less than
1$\%$ from a KS test, so we can say that $blue$ fLSBs are relatively
uniformly distributed. On the other hand, $sequence$ and $red$
fLSBs are different from a uniform spatial distribution with
respective probabilities of 90 and 99$\%$, based on a KS test. In
Fig.~\ref{fig:ad}, the $red$ fLSBs generally tend to be found
  preferentially along the large scale filament of galaxies found by
  Bou\'e et al. (2008). This suggests that $red$ fLSBs could be linked
  with this filament made up of groups infalling toward the Abell~496
  center.

\begin{figure} 
\centering \mbox{\psfig{figure=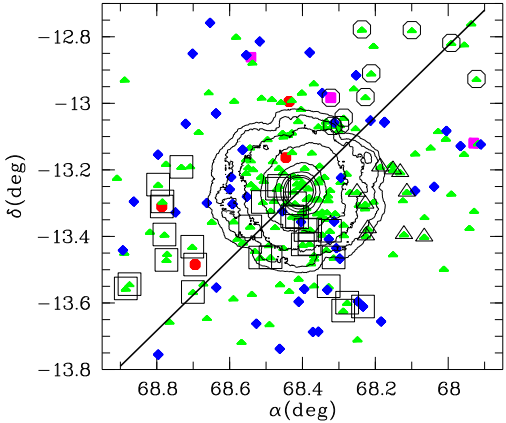,width=6.0truecm,angle=-90}}
\caption[]{Spatial distribution of galaxies of various types with the
  following symbols: red hexagones for ellipticals, purple squares for
  early-type spirals, green triangles for intermediate spirals, and
  blue diamonds for late-type spirals. The spectral types of these
  galaxies were determined in the photo$-z$ computation process fixing
  the redshifts to their spectroscopic values.  The black open symbols
  (squares, triangles and circles) show the three main dynamically
  distinct groups. Contours correspond to the XMM-Newton X-ray
  emission. The black line indicates the direction of the very large
  scale filament of galaxies found by Bou\'e et al. (2008).}
\label{fig:type}
\end{figure}

Since the cluster Abell~496 is believed to be nearly relaxed (Durret
et al. 2000), it is important to determine if it is still experiencing
an infalling activity. We searched for substructures in Abell~496 by
applying the Serna \& Gerbal (1996) method to our large spectroscopic
redshift sample of 596 galaxies (Durret et al. 1999).  
We show in
Fig.~\ref{fig:type} the spatial distribution of galaxies belonging to
various independent dynamical structures inside the Abell~496 cluster.

We only detected three such structures and they are all low mass
structures of a few $10^{12}$~M$_\odot$. These masses are very small
compared to the overall cluster mass of the order of 3.5 $10^{14}$
~M$_\odot$ (e.g. Lagan\'a et al. 2010) and do not prevent us from
classifying the cluster from being relatively well relaxed. The
description of the Serna \& Gerbal method and the full analysis of the
results thus obtained can be found in Appendix~B.

We also see from Fig.~\ref{fig:type} that the two main substructures
are located towards the northwest and southeast of the cluster, that
is roughly along the direction of the large scale filament feeding the
cluster. A third less massive structure is located towards the
west. Cold dark matter hierarchical structure formation models
(e.g. Colberg et al 1999) predict that clusters of galaxies grow via
group accretion. In this context, the cluster substructures
detected along the path of the large scale filament are probably
recent infallen groups.

\begin{figure} 
\centering \mbox{\psfig{figure=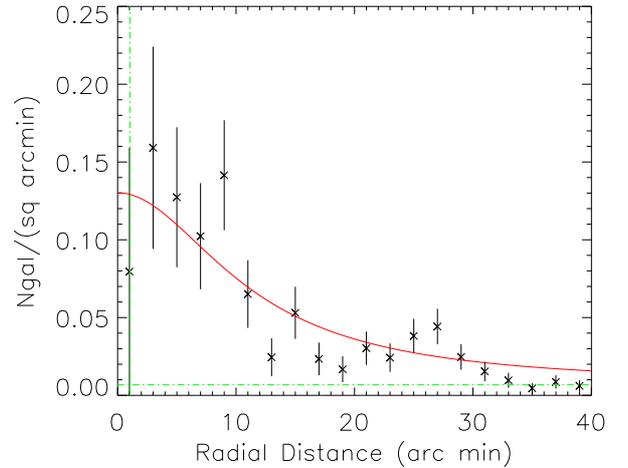,width=7truecm,angle=90}}
\centering \mbox{\psfig{figure=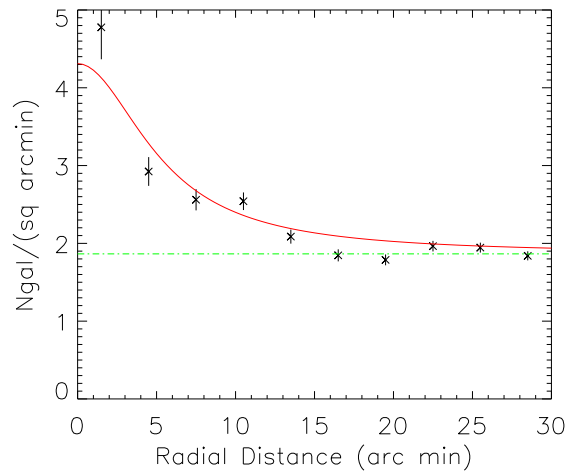,width=7truecm,angle=90}}
\caption[]{Upper figure: number of galaxies per square arcmin versus
  distance to the cluster center, considering the 142 fLSBs with
  photo$-z < 0.2$. The vertical dashed line shows the cluster central
  galaxy radius.  Lower figure: number of galaxies per square arcmin
  versus distance to the cluster center, considering the 5766 galaxies
  with photo$-z < 0.2$ (fLSBs and normal galaxies).  The red lines
  show the best King model fits in both cases. The green horizontal
  lines show the respective background contributions.}
\label{fig:densproflsbs}
\end{figure}

We also quantified the spatial distribution of galaxies with photo$-z
< 0.2$ as a function of radius both for the fLSBs and for the entire
sample of galaxies in our images (including non-fLSB galaxies). For
this, we counted the numbers of galaxies in concentric annuli with
radii varying between 5 and 30~arcmin in steps of 5~arcmin. The
density distributions (number of galaxies per arcmin$^2$) thus
obtained are drawn in Fig.~\ref{fig:densproflsbs}.

These distributions were fit by a King model and we added a
  constant background to take into account non-cluster photo$-z<0.2$
  galaxies:
$$I(r)= P(0) + P(1)/(1+(r/P(2))^2). $$
For the 142 fLSBs with photo$-z<0.2$, the best fit was obtained for
the following parameters:
\noindent
P(0)=(6.8$\pm$6.3)~10$^{-3}$ arcmin$^{-2}$, 
P(1)=0.12$\pm$0.03~arcmin$^{-2}$,
P(2)=11.19$\pm$0.03~arcmin.

For all the galaxies, the corresponding numbers are:
\noindent
P(0)=1.87$\pm 0.09$~arcmin$^{-2}$, 
P(1)=2.45$\pm 0.54$~arcmin$^{-2}$, 
P(2)=5.29$\pm 0.51$~arcmin.

As can be seen in Fig.~\ref{fig:densproflsbs}, fLSBs in Abell~496 are
not uniformly distributed, but are preferentially found toward the
cluster center, except for a decrease of the number of fLSBs per
arcmin$^2$ in the cluster innermost point.  This data point is most
likely low because it is located inside the central galaxy
radius, thus fLSBs in this region would have been missed by our
analysis. Note that although the fLSBS are concentrated toward the
cluster center they are less concentrated than the whole galaxy
population, as discussed in Section~\ref{sec:discu}.

\subsection{Luminosity function of the fLSBs}

\begin{figure} 
\centering \mbox{\psfig{figure=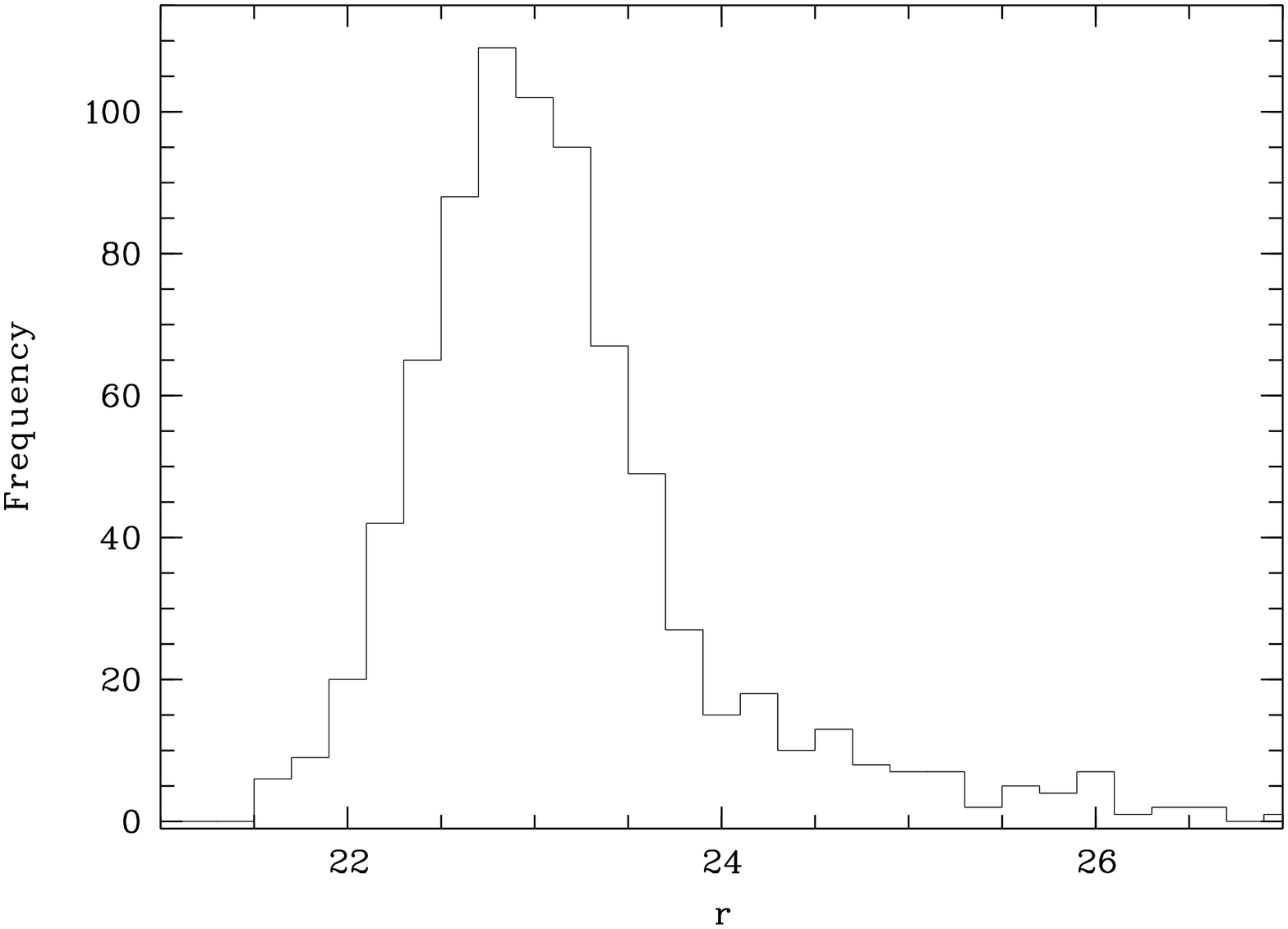,width=6cm,angle=270}}
\centering \mbox{\psfig{figure=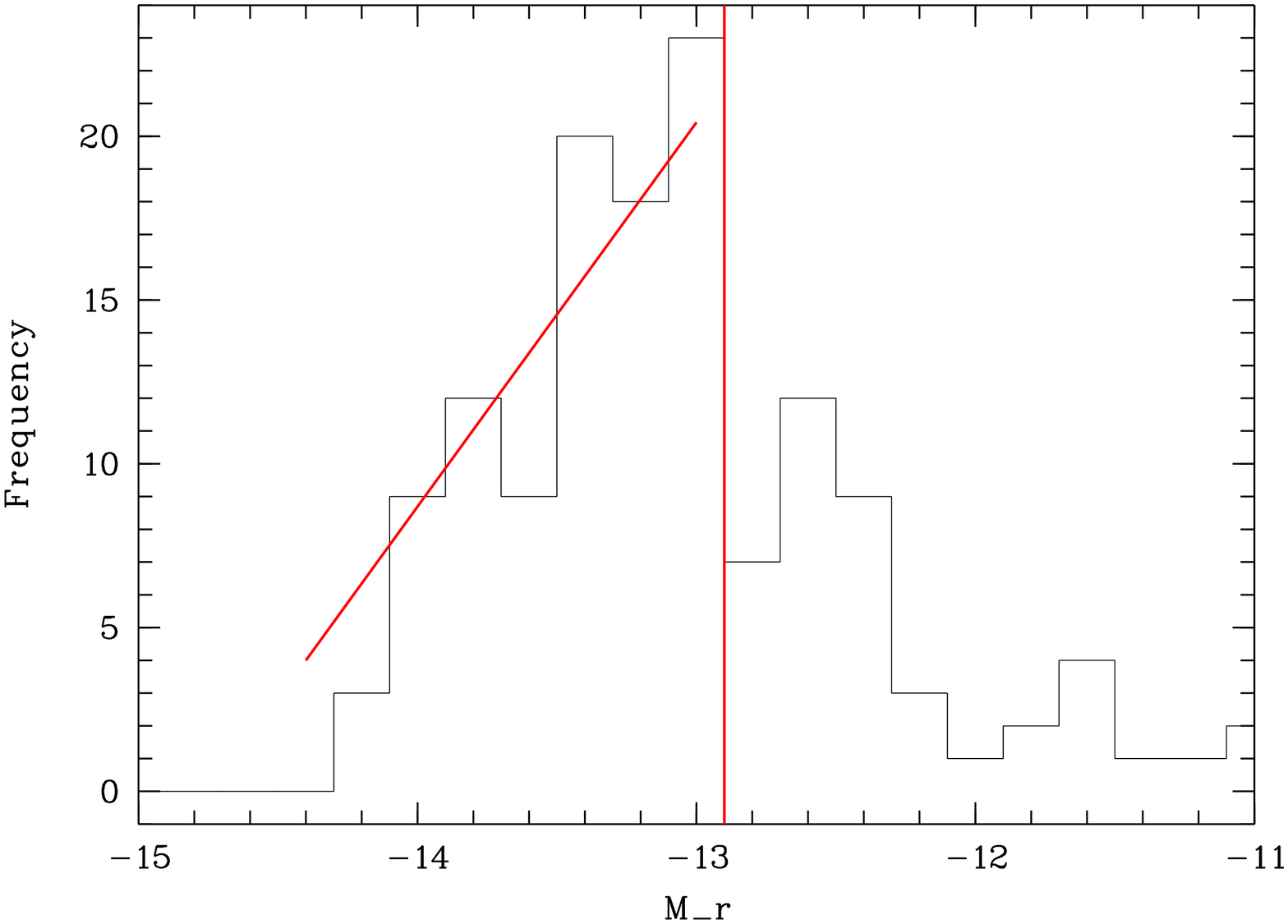,width=6cm,angle=270}}
\caption[]{Upper figure: $r'$-band magnitude histogram of the 783
  fLSBs with available photo$-zs$. Lower figure: luminosity function
  for the 142 fLSBs with photo$-z<0.2$ as a function of absolute
  $r'$-band magnitude (assuming these objects are cluster
  members). The vertical line shows the approximate completeness level
  of the fLSB sample derived from our simulations. The oblique line
  shows the mean slope of the luminosity function for absolute
  $r'$-band magnitude brighter than $-12.9$ (see text).}
\label{fig:histos}
\end{figure}

Before computing a luminosity function for fLSBs, it is important to
investigate the completeness level of our sample. We show in
Fig.~\ref{fig:histos} the magnitude histogram of the 783 fLSBs with
available photo$-z$s and the luminosity function of the 142 fLSBs with
photo$-z<0.2$.  The peak of the magnitude histogram for the 783 fLSBs
is located close to $r'=22.7$. This gives a first estimate of the
completeness limit of the sample. We also performed simulations in
order to have an independent estimate of the completeness in the $r'$
images.

\begin{figure} 
\centering \mbox{\psfig{figure=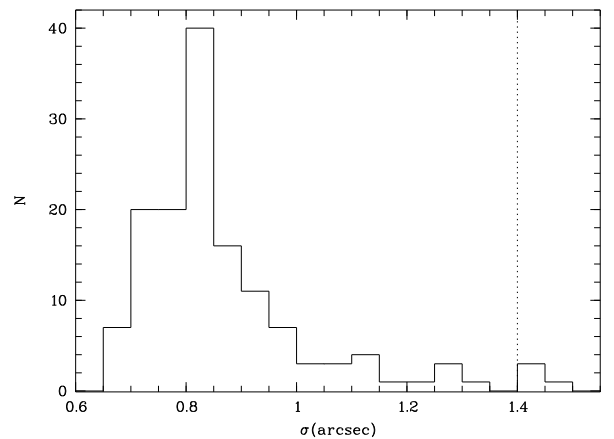,width=6cm,angle=270}}
\caption[]{Dispersion $\sigma$ (in arcsec) of the photo$-z<0.2$
  fLSBs. The dashed vertical line represents the size of our simulated
  fLSBs.}
\label{fig:kr}
\end{figure}

\begin{figure} 
\centering \mbox{\psfig{figure=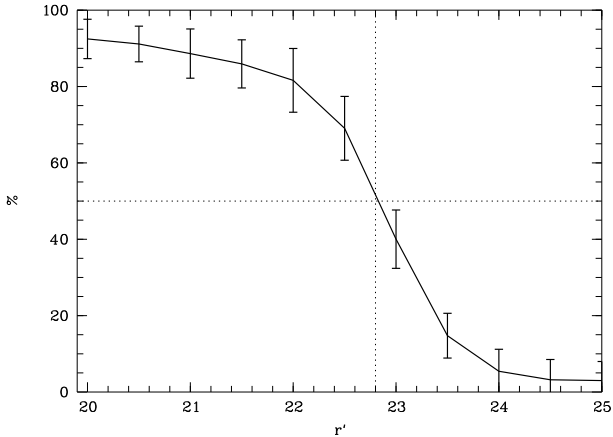,width=6cm,angle=270}}
\caption[]{Percentage of recovered fLSBs in our simulations as a
  function of the $r'$ magnitude.  The dotted lines show the 50$\%$
  completeness level.}
\label{fig:simu}
\end{figure}

The simulation adds artificial objects of different shapes and
magnitudes to the CCD images and then attempts to recover them by
running SExtractor again with the same parameters used for primary
object detection (see Adami et al. 2006 for more details). In this
way, the completeness is measured on the original CCD frames.  We
estimated the completeness of our catalog for fLSBs using simulated
point-like objects with a Gaussian profile of FWHM 3.3 arcsec
($\sigma$=1.4 arcsec). This is the typical maximal size of a fLSB in
our catalog (see Fig.~\ref{fig:kr}). We also divided the full field of
view in 100 different sub-regions to have the completeness at
different locations in the cluster.  The percentage of recovered fLSBs
as a function of the $r'$ magnitude is shown in Fig.~\ref{fig:simu},
where error bars show the variation among these 100 regions. We can
see that we reach a 50\% completeness at $r'\sim$22.8. This estimate
is similar to the value of the peak of the fLSB magnitude histogram.
It also represents an underestimate of the true fLSB completeness
level, as most fLSBs are more compact than a 3.3~arcsec FWHM Gaussian
profile, and therefore easier to detect.

We can see in Fig.~\ref{fig:histos} that the luminosity function
decreases for magnitudes fainter than the completeness limit, as
expected. A power-law fit of the bright part of the luminosity
function gives a mean slope of $-1.2\pm 0.1$. This is significantly
shallower than the global luminosity function of Bou\'e et al. (2008),
who found a slope of $-1.55 \pm 0.05$. This means that the fLSBs
cannot be responsible for all the increase of the global galaxy
luminosity function of the cluster at faint magnitudes. We probably
start missing fLSBs for $r'$ magnitudes fainter than $\sim$22.8. At
the cluster redshift, this translates to M$_{r'}\sim -12.9$.

\section{Discussion}
\label{sec:discu}

As described above, we have found 142 fLSBs in the direction of
Abell~496 with photo$-z<0.2$, out of which about 80\% are probably
cluster members. Their angular density profile is well fit by a King model
with a core radius about twice as large as for normal galaxies. The
King distribution of fLSBs in Abell~496 is very different from what
was observed in Coma by ASU06, where fLSBs do not follow any King-like
distribution. This difference is consistent with the idea that
Abell~496 is relaxed while Coma is not. Furthermore, the wider
radial distribution of the fLSBs versus normal galaxies in
Abell~496 is consistent with the idea that mass segregation has
occurred in Abell~496.

The detected fLSBs fall reasonably well on the extension of the
bright end of the color-magnitude relation established by Bou\'{e}
et al. (2008).  The fact that we have found (see section 3.1) the
$\pm 1\sigma$ interval for the fLSBs around the red sequence similar
in Abell~496 and Coma, one a relaxed cluster, the other not, fits
with the idea that the relaxation state of the cluster does not
influence the position of the fLSBs on the red-sequence.  The
similar red-sequence width in both clusters could be attributed to
$sequence$ fLSBs having evolved in similar groups that fell into the
clusters later, as suggested by ASU06.

On scales of $\geq 1$~Mpc, we note that there is a filament in the
normal galaxy population with redshifts $< 0.2$ found by Bou\'{e} et
al (2008).  The filament extends along a north-west to south-east
line.  In Fig.~4 we can see that $red$ fLSBs (with redshifts $<
0.2$) seem to have an anisotropic distribution similar to the
filament found by Bou\'{e} et al. However, {\it blue} fLSBs show no
obvious anisotropic distribution, suggesting they had a different
evolutionary history.  {\it Blue} fLSBs are perhaps the remnants of
tidally disrupted late-type galaxies as hypothesized by ASU06 for
Coma.

In terms of tidal disruption, we note that the spatial distribution
of fLSBs seems to show no holes in the cluster center, which is not
the case for Coma (ASU06). For Coma the fLSBs could have been
destroyed by tidal disruption due to the massive D galaxies in the
Coma core. In contrast, there is only one central galaxy in the
center of Abell~496, which could produce much less tidal disruption.
It is beyond the scope of this work, though, to carry out numerical
simulations to verify or falsify the idea that fLSBs are tidally
destroyed in the core of Coma and not in Abell~496.

\begin{acknowledgements}

We thank the referee for useful comments.  
We acknowledge our collaboration with G.~Bou\'e and V.~Cayatte
  during the first stages of this project and are grateful to
  T. Lagan\'a for giving us her XMM-Newton images. M.~Ulmer thanks
  UPMC, IAP, Aix-Marseille~I University, and LAM for their hospitality 
  during the different stages of this project, and Bryant Smith and Nicholas
  Logenbaugh for software support.

\end{acknowledgements}

\appendix

\section{Example of postage stamp images}

\begin{figure*} 
\centering \mbox{\psfig{figure=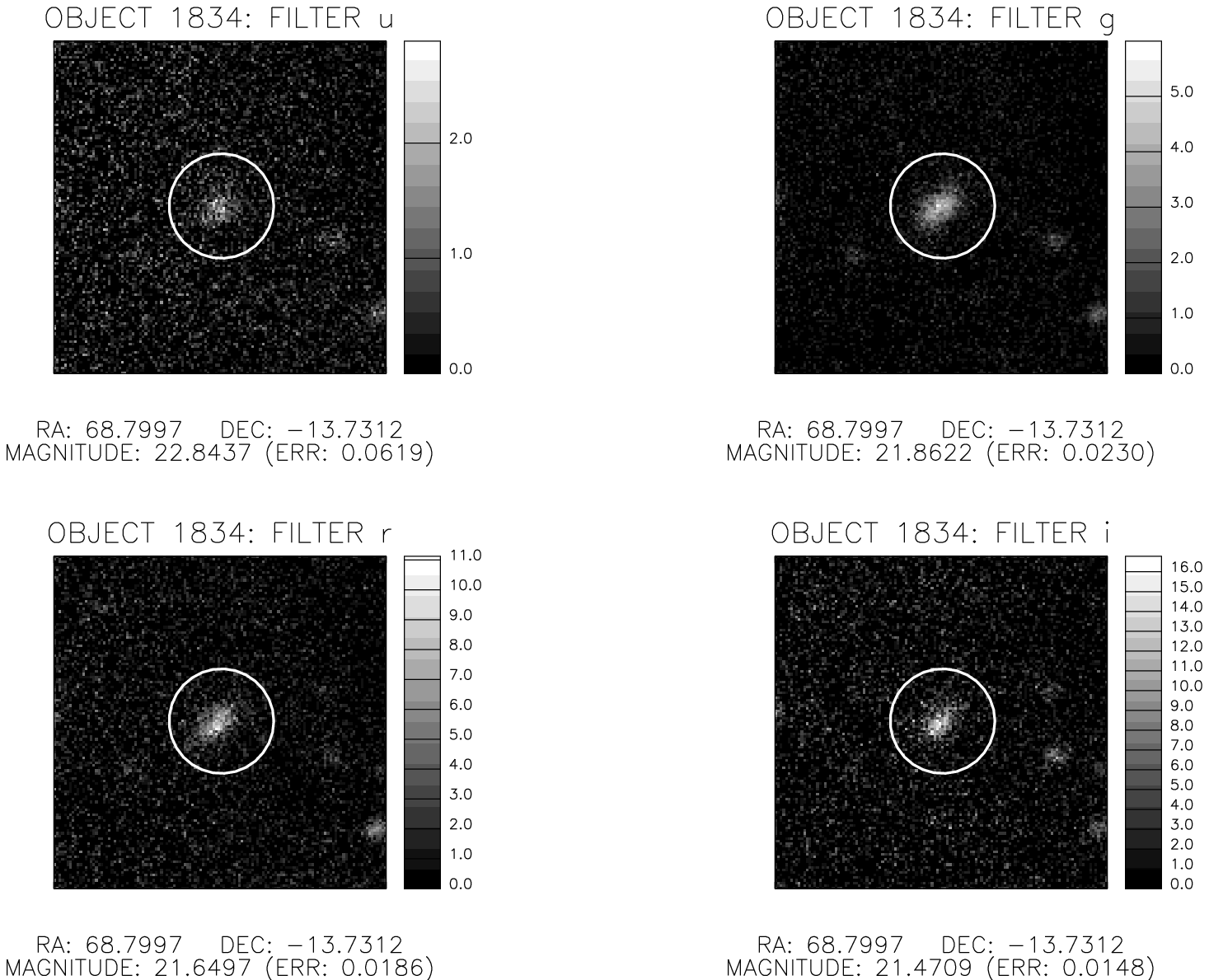,width=13truecm}}
\caption[]{Postage stamp images of one fLSB in the four photometric
  bands (galaxy \#128 in Table~\ref{tab:liste3}). }
\label{fig:post}
\end{figure*}

\begin{figure} 
\centering \mbox{\psfig{figure=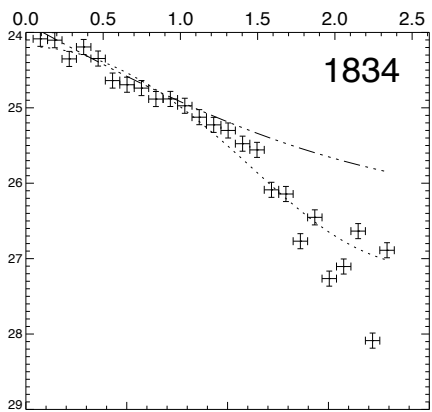,width=6truecm}}
\caption[]{$r'$ band surface brightness profile (in mag~arcsec$^{-2}$
  as a function of radius in arcsec for the galaxy of
  Fig.~\ref{fig:post}). The dot-dashed and dotted lines show the Gaussian
  and exponential fits respectively.}
\label{fig:prof}
\end{figure}

Postage stamps images in the four bands are shown for one of the 142
fLSB candidates with photo$-z<0.2$ in Fig.~\ref{fig:post} (galaxy \#128
in Table~\ref{tab:liste3}). The corresponding surface brightness
profile is given in Fig.~\ref{fig:prof}.

\begin{table*}
  \caption{Properties of the fLSBs with photo$-z<0.2$ (objects \#1 to \#55). The columns are: (1)~Running number, (2) and (3)~Right ascension and declination for equinox J2000.0, (4) to (11) magnitudes in the four bands and corresponding errors, (12)~photometric redshift. Magnitude errors of 88.00 correspond to the fLSBs for which the brightness profile fit by a Gaussian did not converge (see Section~2.3).  }
\begin{tabular}{rrrrrrrrrrrr}
\hline\hline
Nb  &  RA (J2000.0) & DEC (J2000.0) & $u^*$ & err($u^*$) & $g'$ & err($g'$) & $r'$ & err($r'$) & $i'$ & err($i'$) & photo$-z$ \\
\hline
  1 & 67.9070 & -13.0796 & 23.91 &  0.04 & 22.75 &  0.02 & 22.37 &  0.02 & 22.02 &  0.01 & 0.09 \\
  2 & 67.9083 & -12.7808 & 23.26 &  0.03 & 22.59 &  0.02 & 22.50 &  0.02 & 22.62 & 88.00 & 0.19 \\
  3 & 67.9100 & -12.8342 & 23.26 &  0.03 & 22.26 &  0.02 & 22.08 &  0.01 & 22.56 & 88.00 & 0.13 \\
  4 & 67.9101 & -13.0875 & 24.23 &  0.04 & 23.21 &  0.03 & 22.69 &  0.02 & 22.52 &  0.02 & 0.12 \\
  5 & 67.9145 & -13.5382 & 23.60 &  0.04 & 22.79 &  0.03 & 22.45 &  0.02 & 22.51 & 88.00 & 0.02 \\
  6 & 67.9160 & -13.1696 & 26.29 &  0.25 & 25.34 &  0.11 & 25.32 &  0.13 & 27.15 & 88.00 & 0.16 \\
  7 & 67.9191 & -13.7203 & 23.09 &  0.04 & 22.10 &  0.02 & 21.78 &  0.02 & 21.71 &  0.01 & 0.15 \\
  8 & 67.9262 & -12.7923 & 24.34 &  0.05 & 23.34 &  0.03 & 23.14 &  0.03 & 23.04 & 88.00 & 0.11 \\
  9 & 67.9270 & -13.2932 & 23.86 &  0.04 & 23.63 &  0.04 & 24.15 &  0.16 & 24.21 &  0.23 & 0.11 \\
 10 & 67.9505 & -13.1668 & 24.04 &  0.04 & 23.15 &  0.03 & 22.69 &  0.02 & 22.72 &  0.02 & 0.16 \\
 11 & 67.9619 & -13.7119 & 25.21 &  0.18 & 24.83 &  0.14 & 24.13 &  0.09 & 25.27 &  0.15 & 0.10 \\
 12 & 67.9875 & -13.2860 & 24.42 &  0.06 & 23.32 &  0.03 & 22.77 &  0.02 & 22.61 &  0.02 & 0.10 \\
 13 & 67.9895 & -12.9471 & 23.48 &  0.03 & 22.16 &  0.02 & 21.73 &  0.01 & 20.87 & 88.00 & 0.11 \\
 14 & 67.9902 & -13.2069 & 24.75 &  0.07 & 23.76 &  0.04 & 23.40 &  0.03 & 22.83 &  0.02 & 0.11 \\
 15 & 68.0028 & -12.9718 & 24.47 &  0.06 & 23.44 &  0.03 & 23.06 &  0.03 & 23.16 &  0.03 & 0.14 \\
 16 & 68.0050 & -13.1984 & 24.09 &  0.05 & 22.93 &  0.02 & 22.66 &  0.03 & 23.93 &  0.27 & 0.16 \\
 17 & 68.0236 & -12.7699 & 24.40 &  0.05 & 23.05 &  0.02 & 22.54 &  0.02 & 22.48 & 88.00 & 0.12 \\
 18 & 68.0237 & -12.9592 & 24.63 &  0.06 & 23.54 &  0.03 & 23.38 &  0.03 & 23.02 &  0.02 & 0.11 \\
 19 & 68.0283 & -12.8521 & 24.21 &  0.06 & 23.08 &  0.02 & 22.60 &  0.02 & 23.02 & 88.00 & 0.14 \\
 20 & 68.0370 & -12.7976 & 23.70 &  0.04 & 22.77 &  0.02 & 22.53 &  0.02 & 22.46 & 88.00 & 0.16 \\
 21 & 68.0406 & -12.9270 & 24.07 &  0.04 & 22.86 &  0.02 & 22.31 &  0.02 & 21.76 & 88.00 & 0.09 \\
 22 & 68.0476 & -12.8429 & 25.75 &  0.16 & 25.62 &  0.12 & 25.61 &  0.18 & 25.10 & 88.00 & 0.09 \\
 23 & 68.0510 & -13.6045 & 22.37 &  0.02 & 21.76 &  0.02 & 21.57 &  0.01 & 21.55 &  0.01 & 0.14 \\
 24 & 68.0940 & -12.9525 & 25.07 &  0.15 & 24.42 &  0.07 & 24.68 &  0.14 & 23.78 &  0.09 & 0.07 \\
 25 & 68.0952 & -12.8815 & 22.99 &  0.03 & 22.01 &  0.01 & 21.71 &  0.01 & 20.95 & 88.00 & 0.16 \\
 26 & 68.0961 & -12.8760 & 24.68 &  0.06 & 23.73 &  0.03 & 23.11 &  0.02 & 23.35 & 88.00 & 0.12 \\
 27 & 68.1047 & -13.0568 & 24.65 &  0.08 & 23.23 &  0.03 & 22.82 &  0.02 & 22.21 &  0.01 & 0.09 \\
 28 & 68.1090 & -13.3357 & 24.52 &  0.07 & 23.03 &  0.03 & 22.58 &  0.02 & 22.25 &  0.02 & 0.11 \\
 29 & 68.1514 & -12.9756 & 24.39 &  0.05 & 23.24 &  0.02 & 22.64 &  0.02 & 22.43 &  0.02 & 0.14 \\
 30 & 68.1568 & -13.1103 & 24.04 &  0.05 & 23.11 &  0.03 & 22.63 &  0.02 & 21.79 & 88.00 & 0.17 \\
 31 & 68.1588 & -13.3780 & 23.63 &  0.11 & 22.98 &  0.06 & 22.89 &  0.06 & 22.60 &  0.04 & 0.15 \\
 32 & 68.1594 & -13.3791 & 24.55 &  0.08 & 23.41 &  0.04 & 23.02 &  0.03 & 23.08 & 88.00 & 0.15 \\
 33 & 68.1613 & -13.3818 & 23.93 &  0.06 & 22.64 &  0.02 & 22.19 &  0.02 & 21.86 &  0.01 & 0.02 \\
 34 & 68.1754 & -13.4260 & 23.61 &  0.04 & 22.76 &  0.02 & 22.35 &  0.02 & 21.90 & 88.00 & 0.12 \\
 35 & 68.2036 & -13.0714 & 24.18 &  0.05 & 23.11 &  0.03 & 22.65 &  0.02 & 22.46 &  0.02 & 0.16 \\
 36 & 68.2042 & -13.3027 & 24.57 &  0.08 & 23.37 &  0.03 & 22.82 &  0.02 & 22.78 & 88.00 & 0.10 \\
 37 & 68.2249 & -12.8091 & 23.78 &  0.04 & 22.93 &  0.02 & 22.47 &  0.02 & 21.74 & 88.00 & 0.07 \\
 38 & 68.2366 & -12.8454 & 24.16 &  0.06 & 22.77 &  0.02 & 22.26 &  0.02 & 21.26 & 88.00 & 0.13 \\
 39 & 68.2472 & -13.1603 & 24.33 &  0.04 & 24.39 &  0.05 & 24.03 &  0.05 & 23.16 &  0.02 & 0.18 \\
 40 & 68.2507 & -13.2009 & 23.55 &  0.03 & 23.25 &  0.03 & 22.79 &  0.02 & 22.54 &  0.02 & 0.11 \\
 41 & 68.2641 & -13.3386 & 25.03 &  0.14 & 23.36 &  0.04 & 22.90 &  0.04 & 22.54 &  0.02 & 0.11 \\
 42 & 68.2670 & -13.1963 & 24.47 &  0.07 & 23.11 &  0.02 & 22.49 &  0.02 & 22.27 &  0.02 & 0.18 \\
 43 & 68.2774 & -13.3377 & 23.41 &  0.06 & 22.68 &  0.04 & 22.41 &  0.04 & 22.28 &  0.05 & 0.05 \\
 44 & 68.2802 & -13.1144 & 24.54 &  0.06 & 24.90 &  0.10 & 24.12 &  0.06 & 23.81 &  0.07 & 0.10 \\
 45 & 68.2808 & -13.3331 & 24.58 &  0.09 & 23.22 &  0.04 & 22.70 &  0.04 & 22.42 &  0.05 & 0.12 \\
 46 & 68.2887 & -13.3069 & 23.12 &  0.09 & 22.68 &  0.04 & 22.26 &  0.05 & 21.91 &  0.03 & 0.01 \\
 47 & 68.2919 & -13.5850 & 23.96 &  0.07 & 23.23 &  0.06 & 22.68 &  0.05 & 22.57 &  0.03 & 0.16 \\
 48 & 68.2966 & -12.8462 & 23.90 &  0.04 & 22.89 &  0.02 & 22.52 &  0.02 & 22.29 &  0.02 & 0.15 \\
 49 & 68.2972 & -13.1187 & 24.68 &  0.09 & 23.23 &  0.03 & 22.76 &  0.02 & 22.52 &  0.02 & 0.09 \\
 50 & 68.3089 & -13.3587 & 24.74 &  0.07 & 23.47 &  0.03 & 22.92 &  0.02 & 22.64 &  0.02 & 0.11 \\
 51 & 68.3124 & -12.8893 & 24.48 &  0.05 & 23.65 &  0.03 & 23.37 &  0.03 & 23.16 & 88.00 & 0.16 \\
 52 & 68.3162 & -12.9501 & 27.03 &  0.31 & 26.43 &  0.20 & 25.93 &  0.13 & 25.37 &  0.09 & 0.07 \\
 53 & 68.3219 & -13.0564 & 23.55 &  0.03 & 22.59 &  0.02 & 22.02 &  0.01 & 21.77 &  0.01 & 0.12 \\
 54 & 68.3289 & -13.4163 & 24.22 &  0.08 & 23.28 &  0.04 & 23.21 &  0.04 & 23.01 &  0.03 & 0.13 \\
 55 & 68.3331 & -13.4824 & 24.94 &  0.16 & 25.19 &  0.16 & 24.75 &  0.13 & 24.57 &  0.10 & 0.01 \\
\hline
\end{tabular}
\label{tab:liste}
\end{table*}

\begin{table*}
\caption{Same as Table~\ref{tab:liste} for objects \#56 to \#110.}
\label{tab:liste2}
\begin{tabular}{rrrrrrrrrrrr}
\hline\hline
 56 & 68.3343 & -13.0432 & 23.22 &  0.04 & 22.59 &  0.02 & 22.25 &  0.03 & 23.17 & 88.00 & 0.05 \\
 57 & 68.3349 & -13.0424 & 24.89 &  0.08 & 24.73 &  0.05 & 23.87 &  0.03 & 23.56 &  0.03 & 0.05 \\
 58 & 68.3396 & -13.1014 & 23.65 &  0.04 & 22.29 &  0.02 & 21.94 &  0.02 & 21.61 &  0.01 & 0.02 \\
 59 & 68.3412 & -13.1677 & 26.71 &  0.43 & 24.87 &  0.06 & 25.86 &  0.21 & 26.06 & 88.00 & 0.08 \\
 60 & 68.3429 & -13.2528 & 23.62 &  0.04 & 22.43 &  0.02 & 21.88 &  0.02 & 21.17 & 88.00 & 0.16 \\
 61 & 68.3435 & -13.1890 & 23.33 &  0.04 & 22.27 &  0.02 & 21.88 &  0.02 & 20.59 & 88.00 & 0.09 \\
 62 & 68.3508 & -13.5357 & 24.48 &  0.10 & 23.31 &  0.04 & 22.81 &  0.03 & 22.49 & 88.00 & 0.09 \\
 63 & 68.3589 & -13.6966 & 25.05 &  0.26 & 24.44 &  0.13 & 24.02 &  0.10 & 23.61 &  0.05 & 0.09 \\
 64 & 68.3623 & -13.1966 & 25.82 &  0.13 & 24.74 &  0.06 & 24.43 &  0.05 & 23.58 & 88.00 & 0.04 \\
 65 & 68.3642 & -13.3606 & 24.04 &  0.06 & 22.70 &  0.02 & 22.13 &  0.02 & 22.02 &  0.02 & 0.12 \\
 66 & 68.3797 & -12.9096 & 24.20 &  0.06 & 23.23 &  0.03 & 22.72 &  0.02 & 22.19 & 88.00 & 0.17 \\
 67 & 68.3830 & -13.3300 & 24.00 &  0.05 & 22.85 &  0.03 & 22.40 &  0.02 & 22.09 & 88.00 & 0.10 \\
 68 & 68.3896 & -13.3450 & 24.09 &  0.07 & 22.96 &  0.03 & 22.42 &  0.02 & 21.65 & 88.00 & 0.10 \\
 69 & 68.3967 & -13.0617 & 24.55 &  0.07 & 23.39 &  0.03 & 23.06 &  0.03 & 22.55 &  0.02 & 0.11 \\
 70 & 68.4025 & -13.3745 & 23.14 &  0.03 & 23.19 &  0.03 & 22.84 &  0.03 & 14.50 & 88.00 & 0.13 \\
 71 & 68.4032 & -12.9952 & 25.63 &  0.21 & 25.53 &  0.22 & 27.65 &  3.41 & 24.80 &  0.34 & 0.13 \\
 72 & 68.4050 & -12.9890 & 23.65 &  0.03 & 23.22 &  0.03 & 22.64 &  0.02 & 22.13 &  0.01 & 0.18 \\
 73 & 68.4103 & -13.1480 & 23.79 &  0.05 & 22.40 &  0.02 & 21.94 &  0.02 & 21.72 &  0.01 & 0.08 \\
 74 & 68.4251 & -12.8447 & 23.54 &  0.04 & 22.49 &  0.02 & 22.25 &  0.02 & 22.09 &  0.01 & 0.17 \\
 75 & 68.4301 & -13.6645 & 22.84 &  0.04 & 22.04 &  0.02 & 21.76 &  0.02 & 21.57 &  0.01 & 0.19 \\
 76 & 68.4366 & -13.3139 & 23.47 &  0.03 & 22.68 &  0.02 & 22.30 &  0.02 & 21.90 &  0.01 & 0.03 \\
 77 & 68.4389 & -13.1513 & 24.57 &  0.06 & 24.07 &  0.04 & 23.49 &  0.03 & 23.20 &  0.03 & 0.11 \\
 78 & 68.4415 & -13.2490 & 23.77 &  0.04 & 22.88 &  0.02 & 22.36 &  0.02 & 22.07 &  0.01 & 0.12 \\
 79 & 68.4465 & -13.3644 & 24.81 &  0.11 & 23.54 &  0.04 & 23.09 &  0.03 & 21.89 & 88.00 & 0.05 \\
 80 & 68.4475 & -13.6480 & 24.94 &  0.16 & 26.39 &  0.43 & 26.11 &  0.65 & 24.47 &  0.17 & 0.16 \\
 81 & 68.4486 & -13.1831 & 24.30 &  0.07 & 23.07 &  0.03 & 22.45 &  0.02 & 22.90 & 88.00 & 0.18 \\
 82 & 68.4502 & -12.8030 & 24.36 &  0.04 & 23.69 &  0.04 & 23.29 &  0.03 & 22.60 &  0.02 & 0.08 \\
 83 & 68.4568 & -13.2141 & 23.75 &  0.04 & 22.81 &  0.02 & 22.16 &  0.02 & 22.17 & 88.00 & 0.01 \\
 84 & 68.4716 & -13.2397 & 26.34 &  0.45 & 24.87 &  0.11 & 23.57 &  0.04 & 22.32 &  0.02 & 0.06 \\
 85 & 68.4732 & -13.5865 & 23.04 &  0.05 & 21.91 &  0.02 & 21.67 &  0.02 & 21.41 &  0.02 & 0.01 \\
 86 & 68.4996 & -13.4164 & 24.83 &  0.10 & 23.19 &  0.03 & 22.58 &  0.02 & 22.66 &  0.02 & 0.09 \\
 87 & 68.5051 & -13.2755 & 24.31 &  0.07 & 22.86 &  0.03 & 22.38 &  0.02 & 22.05 & 88.00 & 0.13 \\
 88 & 68.5112 & -13.2345 & 23.69 &  0.05 & 22.59 &  0.02 & 22.16 &  0.02 & 21.99 & 88.00 & 0.15 \\
 89 & 68.5113 & -13.1664 & 23.29 &  0.04 & 21.97 &  0.02 & 21.56 &  0.01 & 21.26 &  0.01 & 0.09 \\
 90 & 68.5196 & -13.3452 & 24.48 &  0.09 & 23.14 &  0.03 & 22.75 &  0.03 & 22.45 &  0.02 & 0.09 \\
 91 & 68.5219 & -13.2690 & 23.90 &  0.04 & 23.31 &  0.03 & 22.55 &  0.02 & 22.35 &  0.02 & 0.16 \\
 92 & 68.5226 & -13.1325 & 24.33 &  0.06 & 23.00 &  0.03 & 22.49 &  0.02 & 22.34 &  0.02 & 0.12 \\
 93 & 68.5261 & -13.6989 & 23.24 &  0.06 & 22.29 &  0.03 & 22.03 &  0.02 & 22.25 & 88.00 & 0.08 \\
 94 & 68.5351 & -13.3788 & 23.81 &  0.05 & 23.34 &  0.03 & 23.23 &  0.03 & 23.28 & 88.00 & 0.11 \\
 95 & 68.5359 & -13.1588 & 24.54 &  0.08 & 23.46 &  0.03 & 23.03 &  0.03 & 22.94 &  0.03 & 0.10 \\
 96 & 68.5476 & -13.3125 & 23.59 &  0.05 & 22.36 &  0.02 & 21.97 &  0.02 & 20.84 & 88.00 & 0.09 \\
 97 & 68.5579 & -13.3335 & 24.61 &  0.11 & 23.53 &  0.04 & 23.08 &  0.03 & 22.67 &  0.02 & 0.09 \\
 98 & 68.5614 & -13.1666 & 25.18 &  0.12 & 24.16 &  0.05 & 23.20 &  0.03 & 22.44 &  0.02 & 0.04 \\
 99 & 68.5682 & -13.1921 & 23.85 &  0.05 & 22.71 &  0.02 & 22.05 &  0.01 & 22.02 &  0.02 & 0.09 \\
100 & 68.5697 & -13.0574 & 23.27 &  0.03 & 22.40 &  0.02 & 22.08 &  0.01 & 21.99 &  0.01 & 0.06 \\
101 & 68.5797 & -12.9160 & 24.32 &  0.05 & 23.23 &  0.03 & 22.73 &  0.02 & 22.44 & 88.00 & 0.12 \\
102 & 68.5885 & -13.2804 & 24.30 &  0.07 & 22.87 &  0.03 & 22.38 &  0.02 & 22.24 &  0.02 & 0.10 \\
103 & 68.5979 & -13.6359 & 23.92 &  0.09 & 23.55 &  0.05 & 23.40 &  0.04 & 23.26 &  0.04 & 0.08 \\
104 & 68.6017 & -13.3695 & 24.04 &  0.06 & 22.83 &  0.02 & 22.30 &  0.02 & 22.07 &  0.01 & 0.12 \\
105 & 68.6066 & -13.2390 & 23.47 &  0.03 & 22.66 &  0.02 & 22.21 &  0.02 & 22.03 &  0.02 & 0.12 \\
106 & 68.6121 & -13.7463 & 23.11 &  0.06 & 22.15 &  0.02 & 21.85 &  0.02 & 22.08 & 88.00 & 0.16 \\
107 & 68.6190 & -13.2040 & 24.59 &  0.07 & 23.01 &  0.02 & 22.64 &  0.02 & 22.37 &  0.02 & 0.05 \\
108 & 68.6193 & -13.7032 & 22.97 &  0.04 & 22.10 &  0.02 & 21.77 &  0.02 & 21.66 &  0.02 & 0.01 \\
109 & 68.6264 & -13.4053 & 23.82 &  0.07 & 22.34 &  0.02 & 21.82 &  0.02 & 20.75 & 88.00 & 0.13 \\
110 & 68.6380 & -13.2445 & 24.34 &  0.06 & 23.12 &  0.03 & 22.60 &  0.02 & 22.53 &  0.02 & 0.12 \\
\hline
\end{tabular}
\end{table*}

\begin{table*}
\caption{Same as Table~\ref{tab:liste} for objects \#111 to \#142.}
\label{tab:liste3}
\begin{tabular}{rrrrrrrrrrrr}
\hline\hline
111 & 68.6452 & -13.3425 & 24.70 &  0.09 & 23.22 &  0.03 & 22.73 &  0.02 & 22.63 &  0.02 & 0.16 \\
112 & 68.6595 & -12.8053 & 24.12 &  0.05 & 22.82 &  0.02 & 22.38 &  0.02 & 21.90 &  0.01 & 0.08 \\
113 & 68.6654 & -13.4054 & 23.68 &  0.06 & 22.76 & 88.00 & 21.89 &  0.01 & 21.61 &  0.01 & 0.12 \\
114 & 68.6654 & -13.1666 & 23.35 &  0.03 & 22.63 &  0.02 & 22.49 &  0.02 & 22.48 &  0.02 & 0.11 \\
115 & 68.6853 & -12.7759 & 24.00 &  0.04 & 23.24 &  0.03 & 22.98 &  0.03 & 22.63 &  0.02 & 0.04 \\
116 & 68.6973 & -12.8727 & 23.66 &  0.04 & 22.39 &  0.02 & 21.92 &  0.01 & 22.06 &  0.02 & 0.13 \\
117 & 68.6976 & -13.6447 & 24.22 &  0.09 & 23.88 &  0.06 & 23.74 &  0.05 & 23.35 & 88.00 & 0.08 \\
118 & 68.7007 & -12.8650 & 23.78 &  0.03 & 23.01 &  0.02 & 22.67 &  0.02 & 22.89 & 88.00 & 0.08 \\
119 & 68.7173 & -13.6308 & 23.45 &  0.06 & 22.19 &  0.02 & 21.82 &  0.02 & 21.46 &  0.01 & 0.09 \\
120 & 68.7301 & -13.2592 & 24.52 &  0.07 & 23.67 &  0.03 & 23.04 &  0.02 & 23.10 &  0.02 & 0.18 \\
121 & 68.7319 & -13.5465 & 23.62 &  0.06 & 22.25 & 88.00 & 21.91 &  0.02 & 21.24 & 88.00 & 0.08 \\
122 & 68.7414 & -13.4252 & 23.56 &  0.07 & 21.12 &  0.03 & 21.63 &  0.03 & 21.42 &  0.03 & 0.05 \\
123 & 68.7461 & -13.5330 & 23.59 &  0.07 & 23.24 & 88.00 & 21.95 &  0.02 & 21.98 & 88.00 & 0.09 \\
124 & 68.7473 & -13.4099 & 23.64 &  0.08 & 22.11 &  0.02 & 22.58 &  0.03 & 22.42 &  0.02 & 0.19 \\
125 & 68.7596 & -12.8151 & 24.28 &  0.06 & 22.71 &  0.02 & 22.24 &  0.02 & 21.96 &  0.02 & 0.10 \\
126 & 68.7683 & -13.3438 & 24.15 &  0.08 & 22.86 &  0.04 & 22.34 &  0.04 & 22.11 &  0.02 & 0.16 \\
127 & 68.7936 & -13.2557 & 24.44 &  0.09 & 23.47 &  0.05 & 23.15 &  0.03 & 23.15 &  0.03 & 0.04 \\
128 & 68.7997 & -13.7312 & 22.84 &  0.06 & 21.86 &  0.02 & 21.65 &  0.02 & 21.47 &  0.01 & 0.15 \\
129 & 68.8218 & -13.5805 & 24.11 &  0.08 & 23.44 &  0.05 & 22.72 &  0.03 & 22.87 & 88.00 & 0.10 \\
130 & 68.8253 & -13.4299 & 24.59 &  0.13 & 23.12 &  0.05 & 24.52 &  0.16 & 23.47 &  0.05 & 0.04 \\
131 & 68.8281 & -13.2653 & 25.05 &  0.10 & 23.85 &  0.04 & 23.44 &  0.03 & 23.08 &  0.03 & 0.14 \\
132 & 68.8350 & -13.1813 & 24.79 &  0.08 & 23.76 &  0.04 & 23.37 &  0.03 & 23.02 &  0.02 & 0.18 \\
133 & 68.8449 & -13.0399 & 23.74 &  0.04 & 23.18 &  0.03 & 23.23 &  0.03 & 22.96 &  0.02 & 0.08 \\
134 & 68.8478 & -12.9385 & 23.33 &  0.03 & 22.50 &  0.02 & 22.27 &  0.02 & 22.29 &  0.02 & 0.16 \\
135 & 68.8568 & -13.3383 & 23.98 &  0.06 & 23.03 &  0.03 & 22.44 &  0.02 & 22.29 &  0.02 & 0.15 \\
136 & 68.8589 & -13.6239 & 24.56 &  0.10 & 23.21 &  0.04 & 22.53 &  0.02 & 21.59 & 88.00 & 0.12 \\
137 & 68.8618 & -13.2476 & 24.10 &  0.06 & 23.76 &  0.05 & 23.37 &  0.04 & 23.00 & 88.00 & 0.08 \\
138 & 68.8622 & -12.8247 & 24.21 &  0.05 & 23.26 &  0.03 & 23.01 &  0.02 & 22.66 &  0.02 & 0.11 \\
139 & 68.8629 & -13.7051 & 23.90 &  0.08 & 22.64 &  0.03 & 21.99 &  0.02 & 22.05 & 88.00 & 0.16 \\
140 & 68.8757 & -13.3549 & 23.61 &  0.05 & 23.33 &  0.04 & 23.21 &  0.04 & 23.07 &  0.03 & 0.08 \\
141 & 68.8812 & -13.3283 & 23.49 &  0.04 & 22.62 &  0.02 & 22.28 &  0.02 & 22.29 &  0.02 & 0.15 \\
142 & 68.9077 & -13.1248 & 23.54 &  0.05 & 22.84 &  0.04 & 22.77 &  0.04 & 22.79 &  0.05 & 0.14 \\
\hline
\end{tabular}
\end{table*}

\section{Search for substructures in Abell~496}

Based on the large spectroscopic and photometric catalogues acquired
for Abell~496 (Bou\'e et al. 2008), we have estimated the
spectral type of each galaxy with the Le Phare photometric
redshift software. Galaxies are then assigned a spectral type:
type 1 for ellipticals, type 2 for early type spirals, type 3 for
intermediate type spirals and type 4 for late type spirals.

In order to search for substructures, we applied the Serna \& Gerbal
(1996) software to galaxies with measured spectroscopic redshifts and
magnitudes. This hierarchical method allows to extract galaxy
substructures or groups from a catalogue containing positions,
magnitudes and redshifts, based on the calculation of their relative
(negative) binding energies.  The method gives as output a list of
galaxies belonging to each group, as well as the information on the
binding energy of the group itself, and on the mass of each
substructure, assuming a mass to luminosity ratio (M/L).  We used here
a M/L ratio in the $r'$ band of 200, as previously assumed for the
Coma cluster by Adami et al. (2005), based on the Coma cluster M/L
ratio given by \L okas \& Mamon (2003).

The Serna \& Gerbal analysis shows the existence of three
substructures (also see Section~4). These all have low masses (smaller
than a few $10^{12}$~M$_\odot$) and therefore their existence does not
contradict the overall relaxed structure of the cluster.

If we analyze the morphological type distribution of the galaxies
belonging to these three substructures (also see Fig.~\ref{fig:type}),
we find that only one galaxy is of type 4 (late type spiral),
corresponding to $\sim$1\% of all the galaxies in substructures. If we
estimate the percentage of type 4 galaxies in the cluster (i.e. in the
[0.0229,0.0429] redshift range) that are not included in
substructures, we find a value of 23\%. The difference between these
two values could be interpreted as indicating that late type spirals
tend to avoid substructures and fall individually into the cluster,
while earlier type galaxies fall into the cluster inside groups.

\end{document}